\documentclass[a4paper,12pt]{article}
\pdfoutput=1

\usepackage{amsmath, authblk, a4wide, color, graphicx, appendix, cite, hyperref, bbold, slashed}

\newcommand{\nbrack}[1]{\left(#1\right)}
\newcommand{\sbrack}[1]{\left[#1\right]}

\def\be{\begin{equation}}
\def\ee{\end{equation}}
\def\ba{\begin{eqnarray}}
\def\ea{\end{eqnarray}}

\def\uno{\mbox{1 \kern-.59em {\rm l}}}

\numberwithin{equation}{section}

\begin{document}

\title{\Large{\textbf{A simple technique for combining simplified models and its application to direct stop production}}}
\author[1]{James Barnard\thanks{\texttt{james.barnard@unimelb.edu.au}}}
\author[2]{Ben Farmer\thanks{\texttt{benjamin.farmer@coepp.org.au}}}
\affil[1]{ARC Centre of Excellence for Particle Physics at the Terascale, School of Physics, University of Melbourne, Victoria 3010, Australia}
\affil[2]{ARC Centre of Excellence for Particle Physics at the Terascale, School of Physics, Monash University, Victoria 3800 Australia}
\date{}
\maketitle

\begin{abstract}
\noindent The results of many LHC searches for supersymmetric particles are interpreted using simplified models, in which one fixes the masses and couplings of most sparticles then scans over a few remaining masses of interest.  We present a new technique for combining multiple simplified models (that requires no additional simulation) thereby highlighting the utility and limitations of simplified models in general, and demonstrating a simple way of improving LHC search strategies.  The technique is used to derive limits on the stop mass that are model independent, modulo some reasonably generic assumptions which are quantified precisely.  We find that current ATLAS and CMS results exclude stop masses up to 340 GeV for neutralino masses up to 120 GeV, provided that the total branching ratio into channels other than top-neutralino and bottom-chargino is small, and that there is no mass difference smaller than 10 GeV in the mass spectrum.  In deriving these limits we place upper bounds on the branching ratios for complete stop pair decay processes for many values of the stop, neutralino and chargino masses.  These are available with this paper.
\end{abstract}

\newpage

\section{Introduction}

Much recent attention has been paid to the problem of naturalness in supersymmetry. Supersymmetry (SUSY) remains one of the best motivated candidates for solving the hierarchy problem of the Standard Model, but the non-observation of sparticle production at the Large Hadron Collider (LHC) and the observed Standard Model-like Higgs mass of 125 GeV are forcing us to reassess the tuning present in SUSY models~\cite{Akula:2011jx,Allanach:2012vj,Baer:2012uy,Balazs:2012qc,Barnard:2012au,Gherghetta:2012gb,Baer:2013xua,Evans:2013jna}. Reactions to the LHC observations range from the pessimistic declaration that SUSY is dead, to the optimistic assertion that Run II of the LHC is bound to produce a discovery. Neither statement is true, and yet it remains difficult to extract model independent limits on sparticle masses from the results of LHC searches.

As an attempt to ameliorate this problem, LHC results are frequently interpreted in terms of simplified models, where one assumes a particular sparticle production and decay process, with branching ratios fixed to 100\%. However, each individual model gives an overly optimistic view of the masses excluded by current searches.  Instead, one should carefully combine the information supplied by different simplified models in order to reach a more realistic conclusion.

Currently, the only surefire way to test a more general model is to perform extensive Monte Carlo simulation and compare the expected signal yield with the original ATLAS or CMS data.  The computational expense of this approach has limited most studies to either simple models~\cite{Desai:2011th,Buchmueller:2011ab,Bechtle:2012zk,Fowlie:2012im,Strege:2012bt,Cabrera:2012vu,Buchmueller:2013exa,Kowalska:2013ica,Han:2013kga} or, in the case of more general models, to heuristic approaches that offer useful insights but, strictly speaking, only apply to a particular set of sample points in a large parameter space~\cite{CahillRowley:2012kx}.  Some studies have explored large parameter spaces with, for example, Markov chain Monte Carlo techniques~\cite{Boehm:2013qva}, but the implementation of LHC constraints is often highly simplified.  While recently developed tools \cite{Drees:2013wra, Kraml:2013mwa, Papucci:2014rja} have greatly streamlined the process, it is still interesting to see whether broadly model independent limits can be derived from the simplified models themselves, this being one of the original motivations for using them.

In this paper we introduce a new method for combining information supplied by simplified model interpretations, using direct stop production as an example.  The technique can give almost instantaneous limits on sparticle masses given modest assumptions, and without the need for additional Monte Carlo simulation.  It is therefore of particular interest to model builders, who are typically after approximate limits that are fast and easy to derive, and to the practitioners of global statistical fits, who typically need to evaluate likelihood values at millions of points in a large parameter space (while the technique does not completely replace the need to evaluate likelihood values, it allows the process to be aborted much earlier for strongly excluded points).  We apply our technique to recent LHC results in order to determine model independent limits on the stop mass as a function of the chargino and neutralino masses.  In doing so, we carefully note which remaining assumptions would invalidate our method, and discuss the suitable generalisations that would make it more widely applicable.  We clearly show the limitations of conclusions based purely on a single simplified model, and we also demonstrate that a simple modification to the most popular design of simplified model greatly increases the utility of the approach.

The paper is structured as follows. In section~\ref{sec:simplifiedModelIntro} we give a brief introduction to simplified models, focussing on the case of direct stop production.  In section~\ref{sec:combinationMethod}, we introduce our technique for combining simplified models in the case of a sparticle with only two possible decay modes, before generalising to an arbitrary number of decay modes.  We apply the technique to obtain limits on the stop quark mass in section~\ref{sec:directStop}, and discuss further generalisations in section~\ref{sec:discussion}.  Finally, we present our conclusions in section~\ref{sec:conclusions}.  In deriving limits on the stop mass, we also constrain the branching ratios for each complete stop pair decay process for a wide variety of stop, neutralino and chargino masses.  These constraints are included as additional material with this paper.

\section{Simplified models for direct stop production}
\label{sec:simplifiedModelIntro}

In the context of SUSY results at the LHC, the most common definition of the term ``simplified model'' is a subset of the phenomenological MSSM, in which the masses of all sparticles are fixed, except for a few of interest, and the decay rates of the sparticles are fixed to trivial values, usually a branching ratio of 100\% into a single final state. One then scans over the few remaining sparticles masses and optimises experimental analyses to gain discovery or exclusion power in as much of the reduced parameter space as possible. The results of an LHC search can then be used to place constraints at an arbitrary confidence level on overall cross-sections within the simplified model and, given a highly specific set of assumptions about the bulk of the SUSY parameters, exclusion limits on the sparticle masses themselves.

The hope is that, by choosing a representative selection of simplified models, one can cover the most interesting topologies and gain experimental reach to almost all typical SUSY models. The limit of applicability of simplified model limits is, however, rarely explored, and neither is there a well defined procedure for combining information from different simplified models to obtain sparticle mass limits in more realistic scenarios (though the use of simplified models to obtain limits in GUT scale SUSY models was considered in~\cite{Gutschow:2012pw}). Given the degree to which experimental searches are optimised on these scenarios, one should certainly investigate whether the strategy currently chosen is the best that can be defined.

As an example, consider the pair production of stop quarks, a topic of considerable interest given the role that stops play in the fine tuning of the Higgs boson mass.  Typical direct stop simplified models assume that stops are pair produced, with a production cross-section dependent only on the mass of the stop, and that all stops decay via a single decay mode.  Of course, more than one decay mode is usually accessible to the stops.  In particular, each stop is often able to decay either to a top quark and a neutralino, or to a bottom quark and a chargino.  The standard simplified model approach in this instance is to construct one simplified model where the stop decays \emph{only} to top-neutralino, and one where the stop decays \emph{only} to bottom-chargino, with the chargino subsequently decaying via an on or off-shell $W$-boson.\footnote{This ensures that one does not have to worry about the effect of sleptons on the problem.  It is in fact a conservative approach, given that the effect of a light slepton will be to increase the leptonic branching ratio of the stop, leading to a greater excess of events with leptons that are more easily distinguishable from Standard Model backgrounds.}  A variety of experimental analyses can then be optimised on each model using different final states, e.g.\ 0 leptons, 1 lepton, 2 leptons or various numbers of $b$-jets.

An analysis designed for a particular stop decay process will usually have reduced sensitivity to other processes.  Stop mass limits taken from a single simplified model are therefore almost certainly too strong, given that one often has a branching ratio of less than 100\% into the final state that the simplified model describes.  For example, in the case of analyses optimised for top-neutralino decays, the searches frequently make use of kinematic top reconstruction, thus reducing the acceptance for bottom-chargino decays.  This is only one example of how the approach used on one model will hurt sensitivity to the other.

Instead, a phenomenologist should only use individual simplified models to constrain the overall cross-section for a given process, i.e.\ the product of the stop pair production cross-section times branching ratio for that particular decay mode. Or, if the production cross-section is known, one can constrain branching ratios directly. However, binding limits on the stop mass itself can still be recovered when simplified model analyses exist for all feasible decay modes.  If this is true, one can take each point in the parameter space of interest (e.g.\ each value of stop and neutralino mass) and check whether there is a choice of branching ratios that satisfies all of the simplified model constraints simultaneously.  If such a choice exists, the corresponding mass values cannot be excluded unless assumptions are made about the branching ratios.  On the other hand, if no solution can be found, the corresponding mass values can be excluded \emph{independently} of the branching ratios.

\section{Combining simplified models}
\label{sec:combinationMethod}

Let us suppose that the above decay modes, one with stops decaying to a top quark and a neutralino and the other with stops decaying to a bottom quark and a chargino, are the only two decay modes accessible to the stop. The case of extra decay modes will be considered in the next subsection.  Let us further assume that a simplified model analysis has been performed for the three possible, complete decay channels:
\begin{itemize}
\item 100\% of stop pairs decaying to top-neutralino 
\item 100\% of stop pairs decaying to bottom-chargino
\item 100\% of stop pairs decaying via mixed decays, i.e.\ one stop decaying via each decay mode.
\end{itemize}
The last of these is unphysical; it is distinct from setting the branching ratios for both decay modes to 50\%.  However, its merits have been discussed in ref.~\cite{Graesser:2012qy} (where a dedicated analysis was also proposed), and we will demonstrate that using it is a useful trick that produces significantly improved limits.

Each analysis results in a constraint on the overall cross-section for its decay channel
\be
\sigma(m_{\tilde{t}},\,m_{\tilde{\chi}^0},\,m_{\tilde{\chi}^{\pm}})<\sigma_{\rm ex}(m_{\tilde{t}},\,m_{\tilde{\chi}^0},\,m_{\tilde{\chi}^{\pm}})
\ee
which is a function of the stop, neutralino and chargino masses.  This information is often made publicly available by the experimental collaborations, so no new simulation is required.  The stop pair production cross-section, $\sigma_P$, is a known function of the sparticle masses, so each cross-section constraint can easily be turned into a constraint on the branching ratio for the complete decay channel
\be\label{eq:Bcons}
b(m_{\tilde{t}},\,m_{\tilde{\chi}^0},\,m_{\tilde{\chi}^{\pm}})<\frac{\sigma_{\rm ex}(m_{\tilde{t}},\,m_{\tilde{\chi}^0},\,m_{\tilde{\chi}^{\pm}})}{\sigma_P(m_{\tilde{t}},\,m_{\tilde{\chi}^0},\,m_{\tilde{\chi}^{\pm}})}\equiv B(m_{\tilde{t}},\,m_{\tilde{\chi}^0},\,m_{\tilde{\chi}^{\pm}})
\ee
which is also a function of the stop, neutralino and chargino masses.  Hereafter, this dependence will no longer be made explicit, but should still be assumed.  In analyses that only consider a single simplified model, points for which $B<1$ are excluded.

Note that some additional model dependence can enter proceedings here, as the stop pair production cross-section often depends on additional parameters, such as the gluino mass.  This can be accounted for by simply rescaling the constraints on the branching ratios defined in eq.~\eqref{eq:Bcons}.  If desired, one can even quickly derive separate limits for several different choices of rescaling to see the overall effect of this model dependence.  A similar approach will be used to deal with ``missing'' decay modes shortly.

\begin{figure}[!t]
\begin{center}
\includegraphics[width=0.4\textwidth]{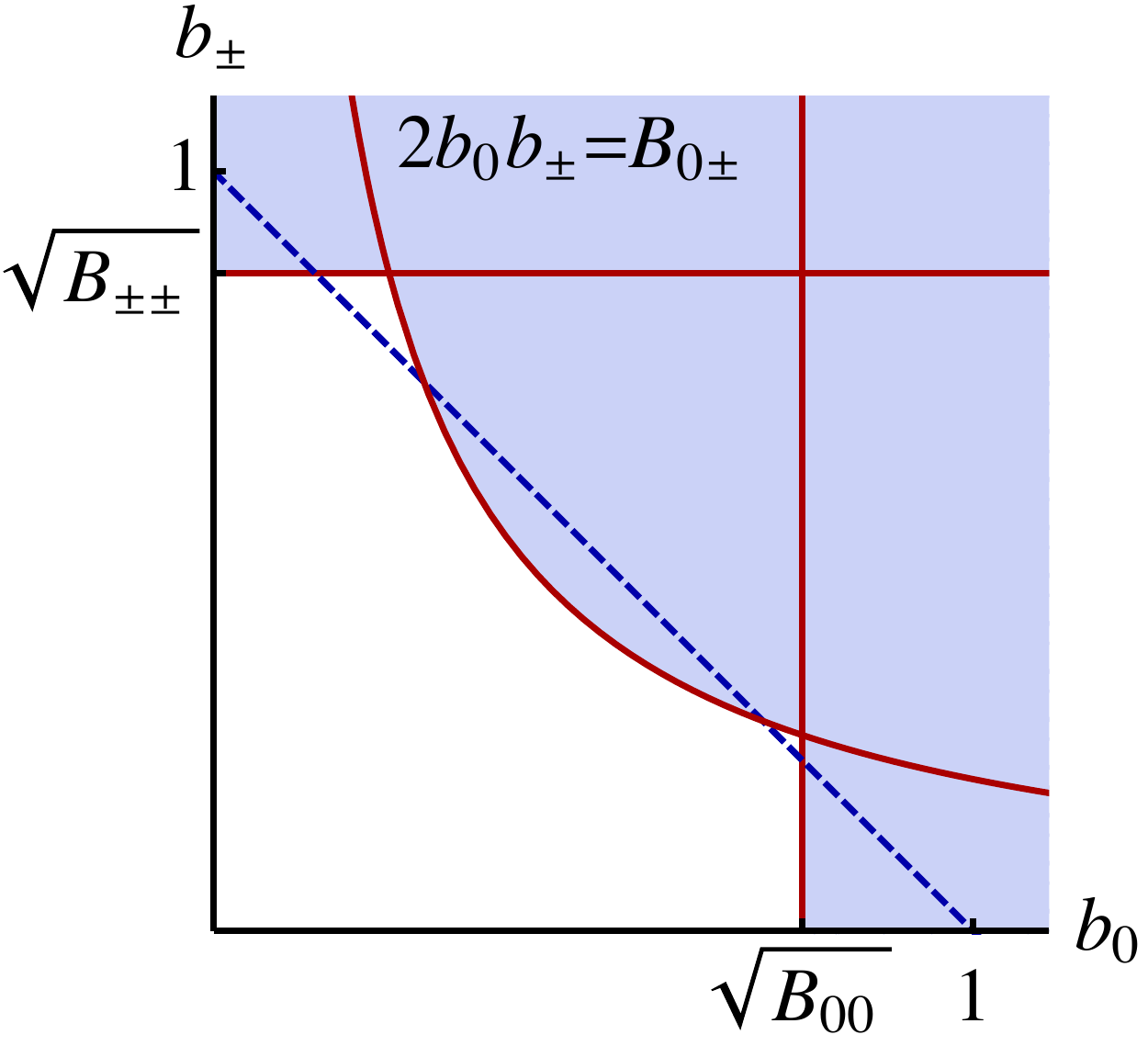}
\end{center}
\caption{How the constraints \eqref{eq:bcons} on complete decay channels (solid, red lines) apply to the branching ratios for each decay mode, $b_0$ and $b_\pm$.  Shaded regions are excluded and the dashed, blue line is the model line $b_0+b_\pm=1$.\label{fig:bb}}
\end{figure}

The branching ratios for each complete decay channel, denoted $b_{00}$, $b_{\pm\pm}$ and $b_{0\pm}$, are related to the branching ratios for the individual stop decay modes, denoted $b_0$ and $b_\pm$, as
\begin{align}
b_{00} & =b_0b_0 &
b_{\pm\pm} & =b_\pm b_\pm &
b_{0\pm} & =2b_0b_\pm
\end{align}
where the subscripts $0$ and $\pm$ denote decays to top-neutralino and bottom-chargino respectively.  We thus have a series of three constraints
\begin{align}\label{eq:bcons}
b_0^2 & <B_{00} & b_\pm^2 & <B_{\pm\pm} & 2b_0b_\pm & <B_{0\pm}
\end{align}
for some functions of the sparticle masses, $B_{00}$, $B_{\pm\pm}$ and $B_{0\pm}$, that can be derived from the simplified model analyses.  These are not the only constraints on $b_0$ and $b_\pm$.  It is also necessary that
\be
b_0+b_\pm=1
\ee
given our earlier assumption that these are the only decay modes available for the stop.  This defines the model line in the $(b_0,b_\pm)$ plane, on which any actual model within this framework must live.  It is helpful to visualise how all of these constraints apply to the decay mode branching ratios using figure \ref{fig:bb}.  A constraint on the branching ratio for a pure decay channel translates directly into a constraint on the branching ratio for the corresponding decay mode, giving a horizontal or vertical line in the $(b_0,b_\pm)$ plane, whereas a constraint on the branching ratio for a mixed decay channel translates into a curve.

The condition that a point on the model line satisfies all of the constraints in eq.~\eqref{eq:bcons} can be written in terms of a single branching ratio, $b_0$ for example, as
\begin{align}
1-\sqrt{B_{\pm\pm}}<{} & b_0<\frac{1}{2}\nbrack{1-\sqrt{1-2B_{0\pm}}} \quad\quad\mbox{or} \nonumber\\
\sqrt{B_{00}}>{} & b_0>\frac{1}{2}\nbrack{1+\sqrt{1-2B_{0\pm}}}.
\end{align}
There are no solutions (i.e.\ both of the above ranges close up) when
\begin{align}\label{eq:bb}
& \sqrt{B_{00}}+\sqrt{B_{\pm\pm}}<1 && \mbox{or} \nonumber\\
& 2\sqrt{B_{00}}-\sqrt{1-2B_{0\pm}}<1 && \mbox{and} && 2\sqrt{B_{\pm\pm}}-\sqrt{1-2B_{0\pm}}<1.
\end{align}
If either of these two conditions are satisfied, there is therefore no way of satisfying all of the constraints \eqref{eq:bcons} simultaneously, and the associated values of stop, neutralino and chargino mass are excluded for all possible configurations of the branching ratios.

One can also use figure \ref{fig:bb} to see when constraints from mixed decays do not contribute.  This occurs when the curve $2b_0b_\pm=B_{0\pm}$ does not cross the model line $b_0+b_\pm=1$ (i.e.\ the constraint on the mixed decay branching ratio does not constrain any realistic models), or when the curve $2b_0b_\pm=B_{0\pm}$ does not enter the box in which $b_0<\sqrt{B_{00}}$ and $b_\pm<\sqrt{B_{\pm\pm}}$ (i.e.\ the constraints on pure decay channels are more powerful).  We conclude that mixed decays do not contribute to any limits if
\begin{align}
B_{0\pm}>\frac{1}{2} && \mbox{or} && B_{0\pm}>2\sqrt{B_{00}B_{\pm\pm}}.
\end{align}
50\% is the maximum possible branching ratio for a mixed decay, so values larger than this should not be considered anyway.

\subsection{More than two decay modes}

The above technique is easily generalised when there are more than two decay modes.  Suppose stops can decay via one of $n$ decay modes, labelled $1,\ldots,n$.  There are then $\frac{1}{2}n(n+1)$ complete decays channels: $n$ in which both stops decay via the same mode, and $\frac{1}{2}n(n-1)$ in which each decays via a different mode.  Denoting the branching ratios for each decay mode as $b_i$, and assuming that a simplified model analysis exists for each complete decay channel, eq.~\eqref{eq:bcons} becomes
\begin{align}\label{eq:bicons}
b_i^2 & <B_{ii} & 2b_ib_j & <B_{ij}\quad(i<j)
\end{align}
and the model line becomes the model plane $\sum b_i=1$.

Considering two of these decay modes, say 1 and 2, in isolation, the above discussion can be recycled, but with each 1 coming from the right hand side of expression for the model line being replaced by $(1-\bar{b}_{12})$, where $\bar{b}_{12}$ is the total branching ratio into decay modes other than 1 and 2
\be
\bar{b}_{ij}=\sum_{k\neq i,j}b_k\quad(i<j).
\ee
In figure \ref{fig:bb} the dashed line is moved towards the origin to intercept each axis at $(1-\bar{b}_{12})$ instead of 1, and the overall result is for eq.~\eqref{eq:bb} to be replaced by
\begin{align}
& \sqrt{B_{11}}+\sqrt{B_{22}}+\bar{b}_{12}<1 && \mbox{or} \\
& 2\sqrt{B_{11}}+\bar{b}_{12}-\sqrt{(1-\bar{b}_{12})^2-2B_{12}}<1 && \mbox{and} && 2\sqrt{B_{22}}+\bar{b}_{12}-\sqrt{(1-\bar{b}_{12})^2-2B_{12}}<1. \nonumber
\end{align}
If either condition is satisfied, there is no way of satisfying the constraints on $b_1$ and $b_2$ simultaneously for the given value of $\bar{b}_{12}$.  If either condition is satisfied for the maximal value of $\bar{b}_{12}$, which eq.~\eqref{eq:bicons} tells us is given by
\be
\bar{B}_{ij}=\sum_{k\neq i,j}\sqrt{B_{kk}}\quad(i<j)
\ee
the constraints on $b_1$ and $b_2$ can never be satisfied simultaneously, hence the model can be excluded for all possible configurations of the branching ratios.  The same reasoning applies to any pair of decay modes, therefore the overall condition for a model to be excluded for all possible configurations of all branching ratios is
\begin{align}
& \sum_{i=1}^n\sqrt{B_{ii}}<1 && \mbox{or} \\
& 2\sqrt{B_{ii}}+\bar{B}_{ij}-\sqrt{(1-\bar{B}_{ij})^2-2B_{ij}}<1 && \mbox{and} && 2\sqrt{B_{jj}}+\bar{B}_{ij}-\sqrt{(1-\bar{B}_{ij})^2-2B_{ij}}<1 \nonumber
\end{align}
for any $i<j$.

One can use the same result if a simplified model analysis does not exist (or is not possible) for some of the pure decay channels, whereupon they can be considered ``missing''.  One can simply add the total branching ratio into missing decay modes to $\bar{B}_{ij}$, i.e.\
\be
\bar{B}_{ij}=b_{\rm miss}+\sum_{k\neq i,j}\sqrt{B_{kk}}\quad(i<j).
\ee
and use the following conditions to exclude points in the parameter space
\begin{align}
& \sum_{i=1}^n\sqrt{B_{ii}}+b_{\rm miss}<1 && \mbox{or} \\
& 2\sqrt{B_{ii}}+\bar{B}_{ij}-\sqrt{(1-\bar{B}_{ij})^2-2B_{ij}}<1 && \mbox{and} && 2\sqrt{B_{jj}}+\bar{B}_{ij}-\sqrt{(1-\bar{B}_{ij})^2-2B_{ij}}<1 \nonumber
\end{align}
for any $i<j$.  Formally, there are no limits on the parameter space in this case as stop pairs could, in principle, decay only via the missing channels, but one can now quickly quantify the dependence of the allowed parameter space on the missing channels by specifying particular values for $b_{\rm miss}$.  It may also be the case that observations from other experiments, or theoretical arguments, strongly suggest a maximum value for $b_{\rm miss}$.

As mentioned earlier, a similar approach can be used to deal with uncertainties in the stop pair-production cross-section.  This quantity appeared in eq.~\eqref{eq:Bcons} when converting constraints on overall process cross-sections into constrains on branching ratios.  If the stop pair-production cross-section were to be scaled by a factor $S$ ($\sigma_P\to S\sigma_P$), one should scale $B\to B/S$ wherever it appears.  Hence the above exclusion conditions become
\begin{align}
& \sum_{i=1}^n\sqrt{B_{ii}}+b_{\rm miss}<\sqrt{S} && \mbox{or} \\
& 2\sqrt{B_{ii}}+\bar{B}_{ij}-\sqrt{(S-\bar{B}_{ij})^2-2B_{ij}}<\sqrt{S} && \mbox{and} && 2\sqrt{B_{jj}}+\bar{B}_{ij}-\sqrt{(S-\bar{B}_{ij})^2-2B_{ij}}<\sqrt{S} \nonumber
\end{align}
for any $i<j$.

\subsection{Comparison with the rigorous combination}

Quite generally, this way of combining simplified models produces weaker limits than rigorous combinations derived on a model-by-model basis.  To get an idea  of how much weaker, we will consider a toy example for the case in which only two decay modes are available to the stops.  Suppose orthogonal analyses (i.e.\ there are no correlations or cross sensitivity) exist for all three complete decay channels, and that each analysis has an expected Standard Model background of five events.  Each analysis observes exactly five events, and has an acceptance-times-efficiency of 50\% for signal events in the decay channel it was designed for only.

Using this information, one can construct the full combined likelihood for the number of stop pairs produced, assuming various, fixed values for the branching ratios.  Applying the standard $CL_s$ limit setting procedure, we find 95\% confidence level limits on the number of stop pairs produced of
\be
\begin{array}{l|rrrrr}
b_0 & 0 & 0.1 & 0.2 & 0.3 & 0.5 \\\hline
\mbox{max stop pairs} & 10 & 11 & 12 & 13 & 13 \\
\sigma_{\rm ex}\mbox{ [pb]} & 0.011 & 0.012 & 0.013 & 0.014 & 0.014 \\
m_{\tilde{t}}\mbox{ limit [GeV]} & 670 & 660 & 650 & 645 & 645
\end{array}
\ee
(for $b_0>0.5$ the results repeat as our toy example is symmetric).  To derive the limits on the stop pair-production cross-section we have assumed that our toy example used 1 fb$^{-1}$ of LHC data, and to subsequently derive the limits on the stop mass we have assumed that only a single stop is light.

The limit of 670 GeV for $b_0=0$ only considers a single complete decay channel, so is equivalent to the limit derived from a single simplified model.  As expected, this naive limit is generally too strong.  If we instead combine the simplified models using the technique described in this section, we find a more realistic but weaker limit of around 580 GeV\@.  While there is no formal confidence level that can be applied to this limit, we can say that  a stop mass of 580 GeV is excluded for all possible values of $b_0$ at a 95\% confidence level by at least one of the individual constraints in eq.~\eqref{eq:bcons}.  This limit is also weaker than the rigorous limit for any given value of $b_0$, as the rigorous limit is able to make use of more information.  Specifically, the full combined likelihood gets an extra suppression if no signal is seen across several independent channels.

It is also interesting to consider the effect of mixed decays in this toy example.  If we neglect the analysis for such decays, the rigorous 95\% confidence level limits weaken to
\be
\begin{array}{l|rrrrrr}
b_0 & 0 & 0.1 & 0.2 & 0.3 & 0.4 & 0.5 \\\hline
\mbox{max stop pairs} & 10 & 13 & 17 & 21 & 25 & 27 \\
\sigma_{\rm ex}\mbox{ [pb]} & 0.011 & 0.014 & 0.018 & 0.022 & 0.026 & 0.028 \\
m_{\tilde{t}}\mbox{ limit [GeV]} & 670 & 645 & 620 & 605 & 580 & 590
\end{array}
\ee
and the limit found by combining simplified models using the technique described in this section weakens to 525 GeV\@.  Mixed decays are clearly important for our toy example, and we find that they remain so when setting actual limits later on.

\subsection{Cross sensitivity and correlations}

So far we have implicitly assumed that the results being combined are independent of one another. This is not always true, although correlations between simplified model analyses generally act to strengthen limits on sparticle masses, unless the analyses are in conflict (e.g.\ if large excesses in one search compete against deficiencies in another).

One way that analyses can overlap is cross sensitivity.  While a given analysis is usually optimised for a given decay channel, it may also have sensitivity to others; even if both stops decay to top-neutralino, for example, a signal event may still be seen in the bottom-chargino analysis. Cross sensitivity between analyses therefore increases the probability of seeing a signal event for every stop pair produced. The overall effect is that a realistic model predicts more signal events than a combination of independent simplified models with appropriate values for the branching ratios. In most cases, one can therefore impose tighter limits on the realistic model.  Note that this is in addition to the effect described in the previous subsection.

If, however, the analyses drift too far from the median background and come into tension with each other (say if an excess is observed in one analysis but not in others) then the limit derived from the rigorous combination can be weakened, possibly below the union of the limits from individual analyses. In this case our simple limits may be too strong for certain choices of the branching ratios. Usually the deviation needs to be fairly large before this occurs, say around $2\sigma$, but it should be kept in mind. This effect can occur even if there are no correlations between the simplified model analyses.

If correlations between analyses become large, i.e.\ there is significant event overlap,\footnote{Correlations can of course come from other sources, for instance the integrated luminosity, but these tend to be small effects} then the above effect can be either amplified or decreased, depending on observations.  For example, if two correlated analyses both see an excess, its combined significance will be less than if the analyses were independent, so the rigorous limit will be less severely weakened than in the uncorrelated case.  Conversely, if one analysis sees an excess and the other a deficiency, then the conflict is greater than in the uncorrelated case, and the rigorous limit will be more severely weakened.

With the above considerations in mind, we see that whether an analysis is inclusive or exclusive is not directly an issue; what is important are the correlations between analyses being combined and the size of any excesses or deficiencies observed.  In general these effects depend on many details of each analysis, so it is less risky to estimate combinations for independent analyses.  The easiest way to keep our simple limits conservative is by using analyses which are as independent as possible, and which do not observe excesses or deficiencies much larger than around $1\sigma$.

In ref.~\cite{Buchmueller:2013exa}, a similar approach to finding model independent limits on the stop mass is followed.  There, the combination is done by constructing a model of the likelihood for each analysis and multiplying them together, that is, making the assumption that the analyses are statistically independent.  This allows competing excesses and deficiencies to be accounted for in each search (which cannot be done with our simple method).  The required statistical independence is most easily achieved with exclusive analyses, since a set of inclusive analyses, depending on the details, has more chance for event overlap with other analyses. So, in this sense, our method relies on similar assumptions about the analyses and is most easily satisfied by a set of exclusive analyses which cover a full set of simplified models, i.e. a set that constrains all possible branching ratios.

Finally, we point out that fully accounting for cross sensitivity and correlations increases the complexity and computational cost of our combination technique, and often requires additional simulation, as the information made publicly available by the experimental collaborations is insufficient for a full likelihood calculation. Hence we will stick with the generally conservative limits produced by the simple technique described previously. To perform the rigorous combination, one would need the acceptance-times-efficiency numbers found by applying each analysis to each simplified model, whereupon a full combined likelihood could be constructed and compared with the data. For a given choice of sparticle masses, this would be a function of the branching ratios only, so one could choose the branching ratios that minimise the likelihood function at every point in the parameter space to derive conservative but rigorous limits on the sparticle masses in the spirit of those we will now present.

\section{Direct stop production}
\label{sec:directStop}

Having developed our combination technique, it is now straightforward to apply it to direct stop production at the LHC\@. We start by considering the data released by the CMS collaboration in ref.~\cite{Chatrchyan:2013xna}.  This analysis sets limits on the stop pair production times branching ratio for pure decays (both stops decaying to top-neutralino or both decaying to bottom-chargino) for a simplified model consisting of a single, unpolarised stop, a single neutralino, and a single chargino with mass
\be
m_{\tilde{\chi}^\pm}=xm_{\tilde{t}}+(1-x)m_{\tilde{\chi}^0}
\ee
for $x=0.75$, 0.5 and 0.25.  Because mixed decays were not considered, only the first condition in eq.~\eqref{eq:bb} can be used to test which points in the parameter space are excluded.

\begin{figure}[!t]
\includegraphics[width=0.48\textwidth]{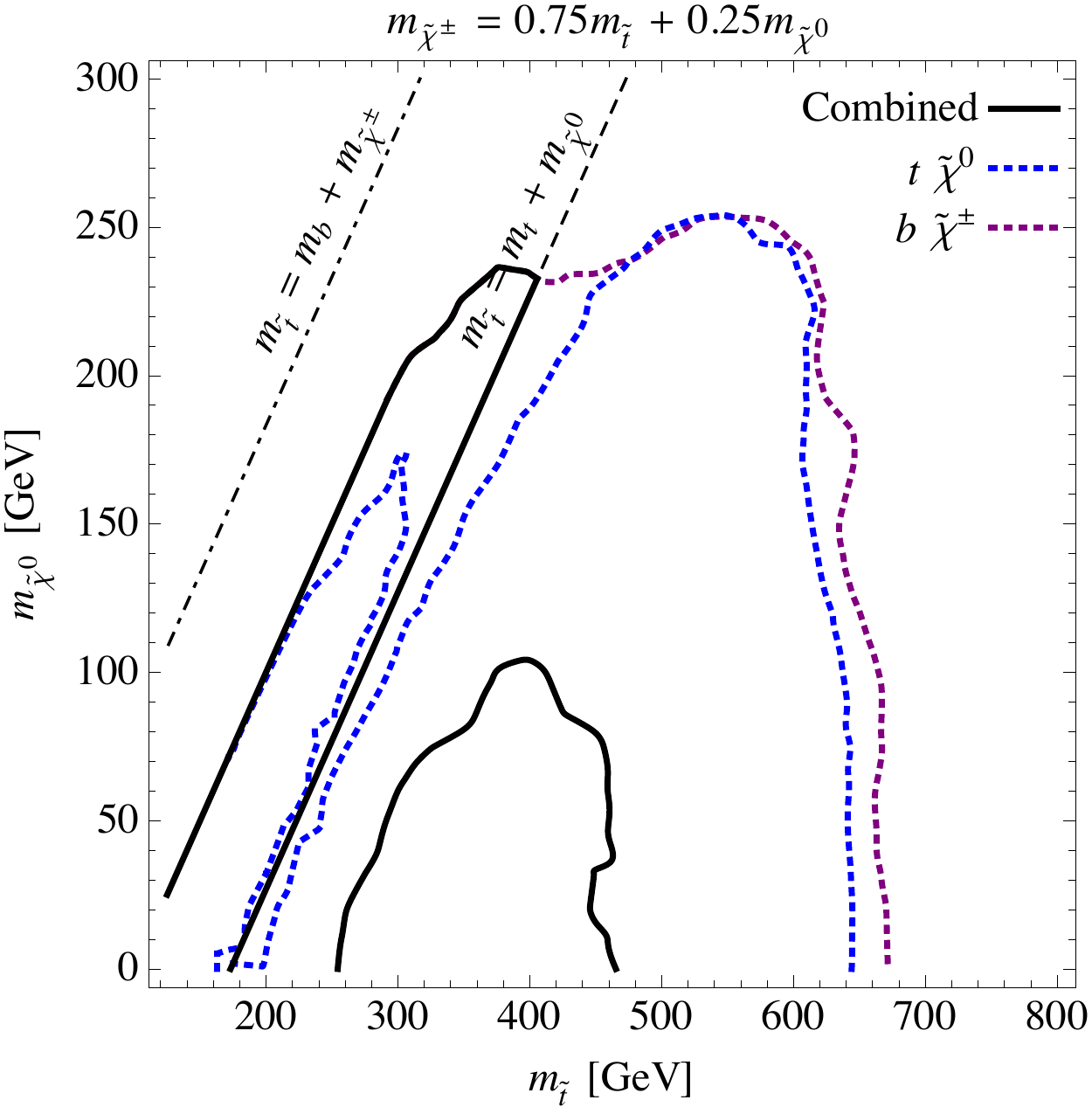}\hspace{5mm}
\includegraphics[width=0.48\textwidth]{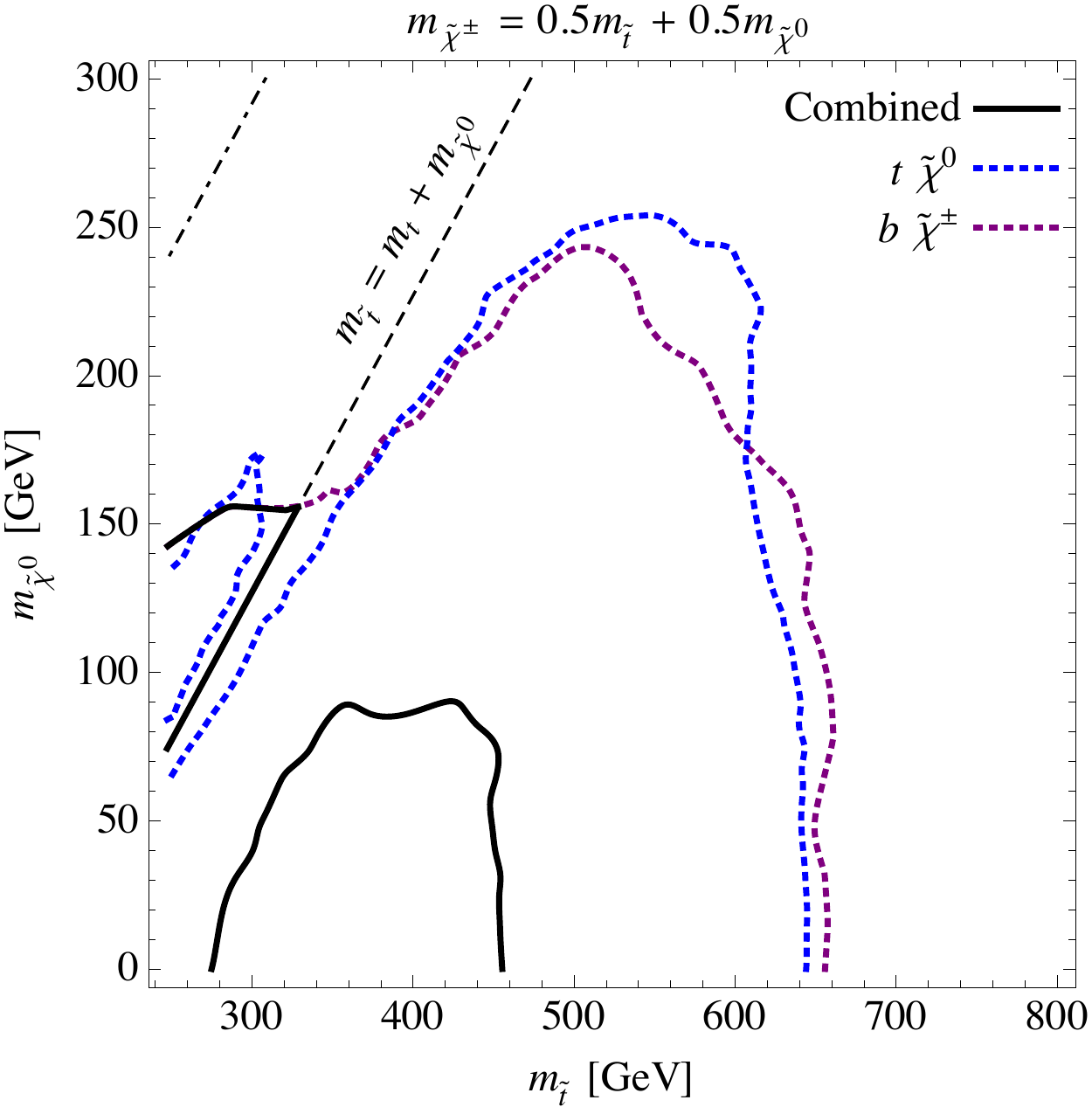}
\caption{A combination of the 95\% $CL_s$ limits released by the CMS collaboration in ref.~\cite{Chatrchyan:2013xna}.  The original limits on the stop and neutralino masses, where a branching ratio of 100\% was assumed for stops decaying either to top-neutralino or bottom-chargino, are shown as dotted contours.  The combined limits, valid for all possible choices of branching ratios, are shown as solid contours.  Ref.~\cite{Chatrchyan:2013xna} also considered a chargino mass $m_{\tilde{\chi}^\pm}=0.25m_{\tilde{t}}+0.75m_{\tilde{\chi}^0}$, but the limits in this case are weaker and none remain valid for all possible choices of branching ratios.\label{fig:CMS}}
\end{figure}

The results of the combination are shown in figure \ref{fig:CMS}.  Demanding that limits on the stop and neutralino masses are valid for all possible choices of branching ratios clearly weakens them.  Even so, the combination of simplified models still produces non-trivial limits for $x=0.75$ and $x=0.5$.  For stop masses lower than $m_t+m_{\tilde{\chi}^0}$ it is reasonable to assume that the branching ratio for bottom-chargino decays is actually 100\%, as these decays remain on-shell while top-neutralino decays go off-shell.  Hence the combined limits coincide with the limits derived from pure bottom-chargino decays.  The opposite would be true in regions where bottom-chargino decays go off-shell but top-neutralino decays remain on-shell.

In addition to the results released by the CMS collaboration, we have applied our technique to the limits on direct stop production released by the ATLAS collaboration \cite{ATLAS-CONF-2013-024, ATLAS-CONF-2013-037, ATLAS-CONF-2013-048, Aad:2013ija}.  Various simplified models were used in these analyses, all of them consisting of a single, unpolarised stop, a single neutralino, and a single chargino, with different assumptions made for the chargino mass.  Again, only pure decays were considered by the collaboration, but this time we have performed additional simulation for mixed decays, where one stop decays via top-neutralino and the other via bottom-chargino, and applied the analyses of refs.~\cite{ATLAS-CONF-2013-024, ATLAS-CONF-2013-037, ATLAS-CONF-2013-048, Aad:2013ija} to them. This enables us to assess whether adding an extra simplified model for mixed decays would improve the information one can extract from the combination of simplified models for pure decays alone.  As an added benefit, we have applied all of the released analyses to the simplified models for pure decays, and we find that this can strengthen the limits in some cases.

\begin{figure}[!t]
\includegraphics[width=0.48\textwidth]{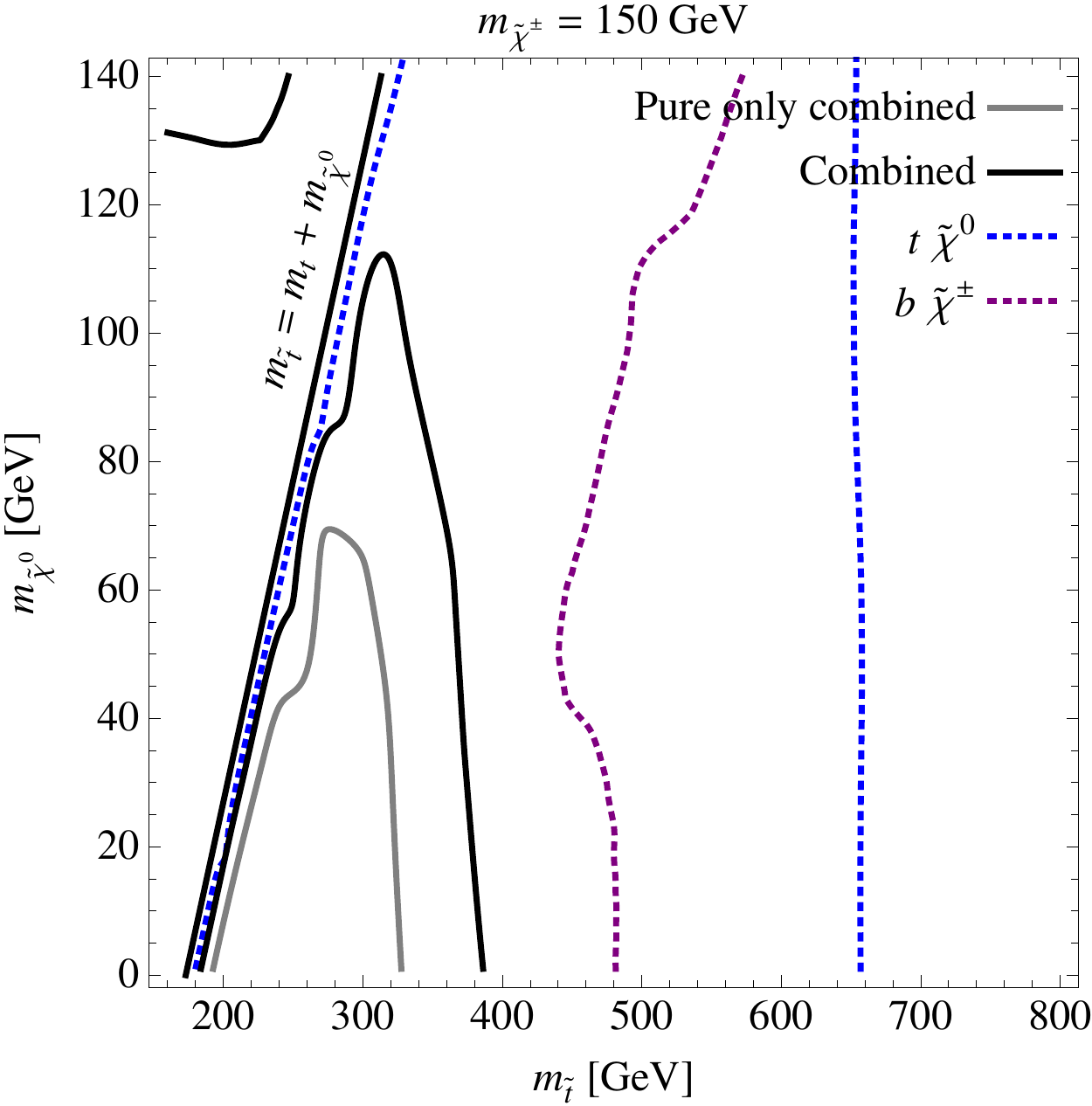}\hspace{5mm}
\includegraphics[width=0.48\textwidth]{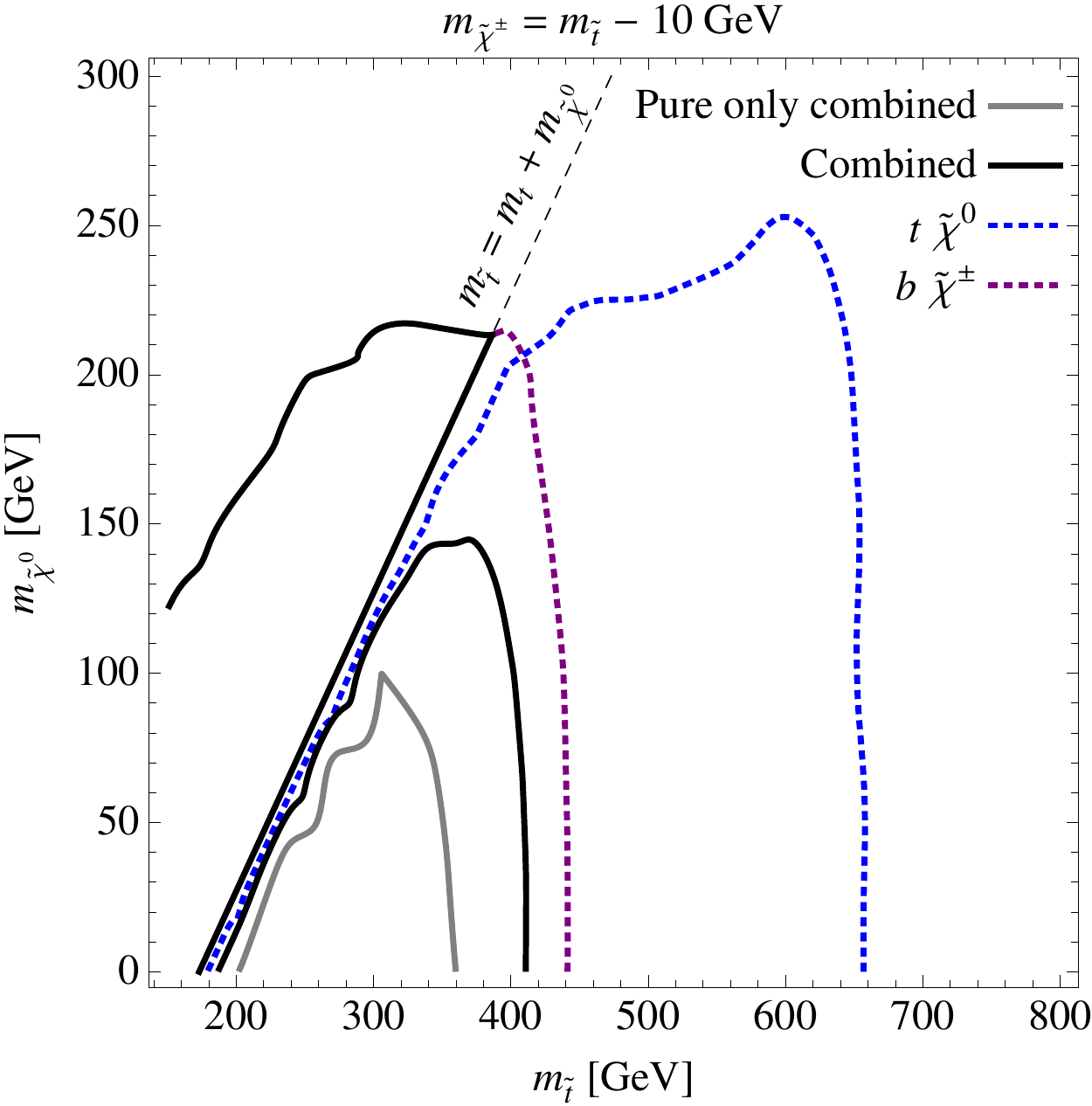}\\
\includegraphics[width=0.48\textwidth]{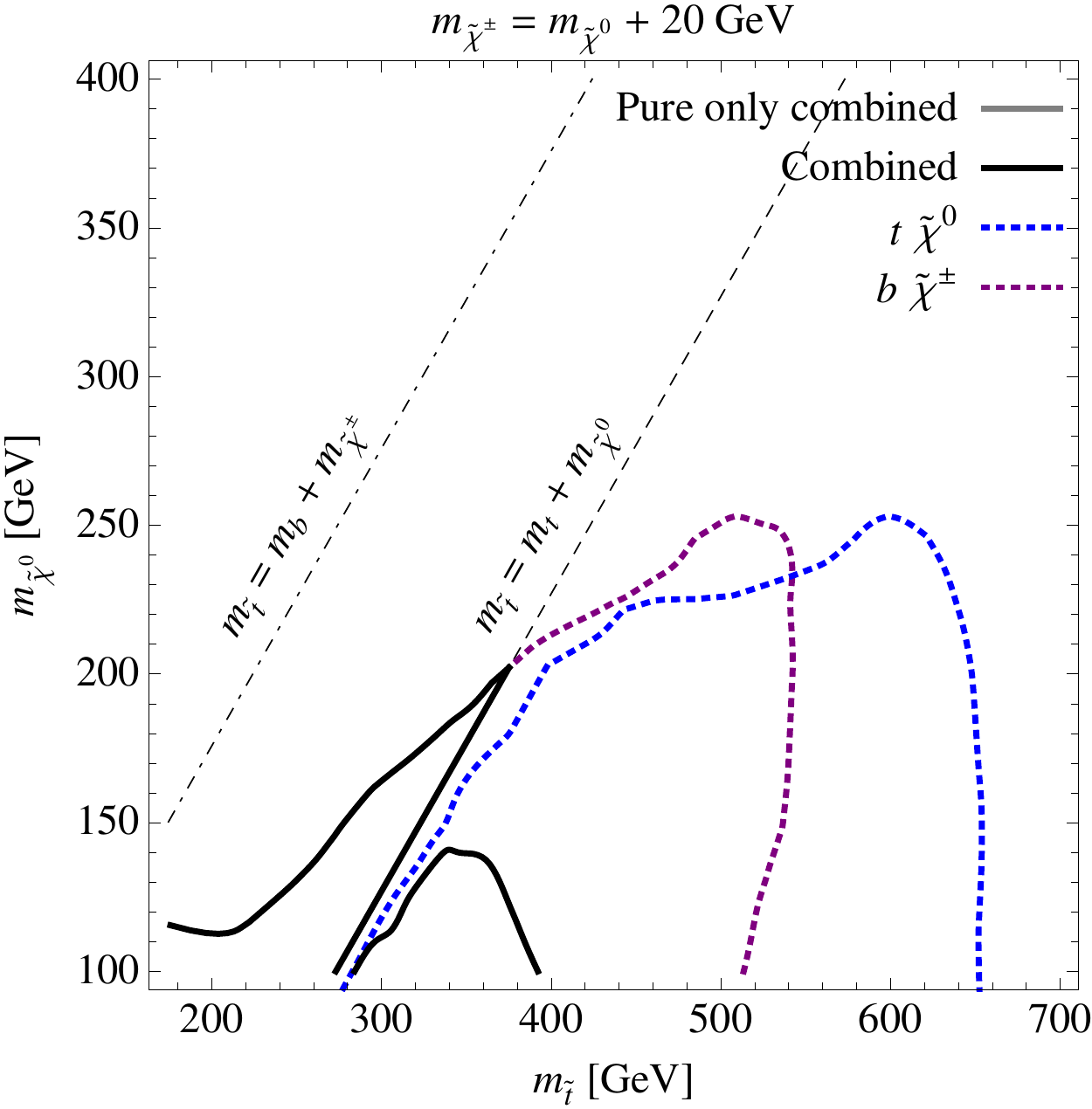}
\caption{A combination of the 95\% $CL_s$ limits released by the ATLAS collaboration in refs.~\cite{ATLAS-CONF-2013-024, ATLAS-CONF-2013-037, ATLAS-CONF-2013-048, Aad:2013ija}.  The original limits on the stop and neutralino masses, where a branching ratio of 100\% was assumed for stops decaying either to top-neutralino or bottom-chargino, are shown as dotted contours.  The combined limits, valid for all possible choices of branching ratios, are shown as solid contours, both with and without constraints coming from mixed decays.  The limits presented here apply all analyses to each simplified model, so are stronger than those already released by the ATLAS collaboration in some cases.\label{fig:ATLAS}}
\end{figure}

Results of the combination are shown in figure \ref{fig:ATLAS}, where combined limits that only use the ATLAS simplified models for pure decays are shown in grey.  These are the limits that can be obtained without performing additional simulation for mixed decays.  Again, we see that the combined limits are significantly weaker than those found using individual simplified models, but still remain non-trivial in some cases.  No grey line is visible for a chargino mass of $m_{\tilde{\chi}^\pm}=m_{\tilde{\chi}^0}+20$~GeV because, in this case, there are no combined limits from pure decays alone when top-neutralino and bottom-chargino decays are both on-shell.

The same figure shows the effect of including a simplified model with mixed decays, with the resulting combination of simplified models shown in black.  In all cases, we see that adding the extra simplified model significantly improves the limits one can obtain by combining simplified models, proving that mixed decays are important and strongly motivating the use of these extra models by the LHC experimental collaborations.  This is true even if the only extra information is that the existing searches are interpreted in the new mixed model. Should a new analysis be optimised on the mixed model, such as that proposed in ref.~\cite{Graesser:2012qy}, the effect will be even greater.

Note that it is important to distinguish the limits that we get from combining simplified models from the true limits obtainable from the LHC data.  Each experimental analysis will have \emph{some} sensitivity to models that are not the same as the model it was optimised on, but detailed Monte Carlo simulation is required to obtain the precise sensitivity of a given analysis to a particular SUSY model.  Our limits represent the information that one can quickly extract from simplified model results \emph{without} further simulation, which is not the same thing, but is instead a very quick, conservative limit.\footnote{While additional simulation is actually performed here, it is only necessary to see the effect of including mixed decays.  If simplified model analyses for mixed decays already existed, as will likely be the case in the future, it would have been straightforward and very quick to include them in the combined limits.}  If a point is excluded by our technique, it can be excluded very quickly and easily (in essence, the technique continuously picks out the worst possible choice of branching ratios for each point in parameter space).  If a point is not excluded by our technique, it is worth further investigation to see if it is actually excluded or not.  For a large scale global statistical fit with millions of likelihood evaluations, our technique provides the ability to reject many points without Monte Carlo simulation, saving a large amount of CPU time.

\subsection{``Model independent'' limits on the stop quark mass}

To close the discussion in this paper, we will apply our technique to derive limits on the stop mass that are conservative, but as model independent as possible, for different assumptions on the branching ratio to missing final states.  Here ``missing'' means a final state for which there is no simplified model information available, not necessarily one that cannot be detected.  While additional simulation is not required for combining existing simplified models, we would like to consider more general chargino masses, for which a comprehensive set of simplified models is not currently available.  Hence we perform simulation to plug the gaps, using simplified models for pure and mixed decays and the ATLAS analyses released in refs.~\cite{ATLAS-CONF-2013-024, ATLAS-CONF-2013-037, ATLAS-CONF-2013-048, Aad:2013ija}.  Details are provided in the appendix.

\begin{figure}[!t]
\begin{center}
\includegraphics[width=0.48\textwidth]{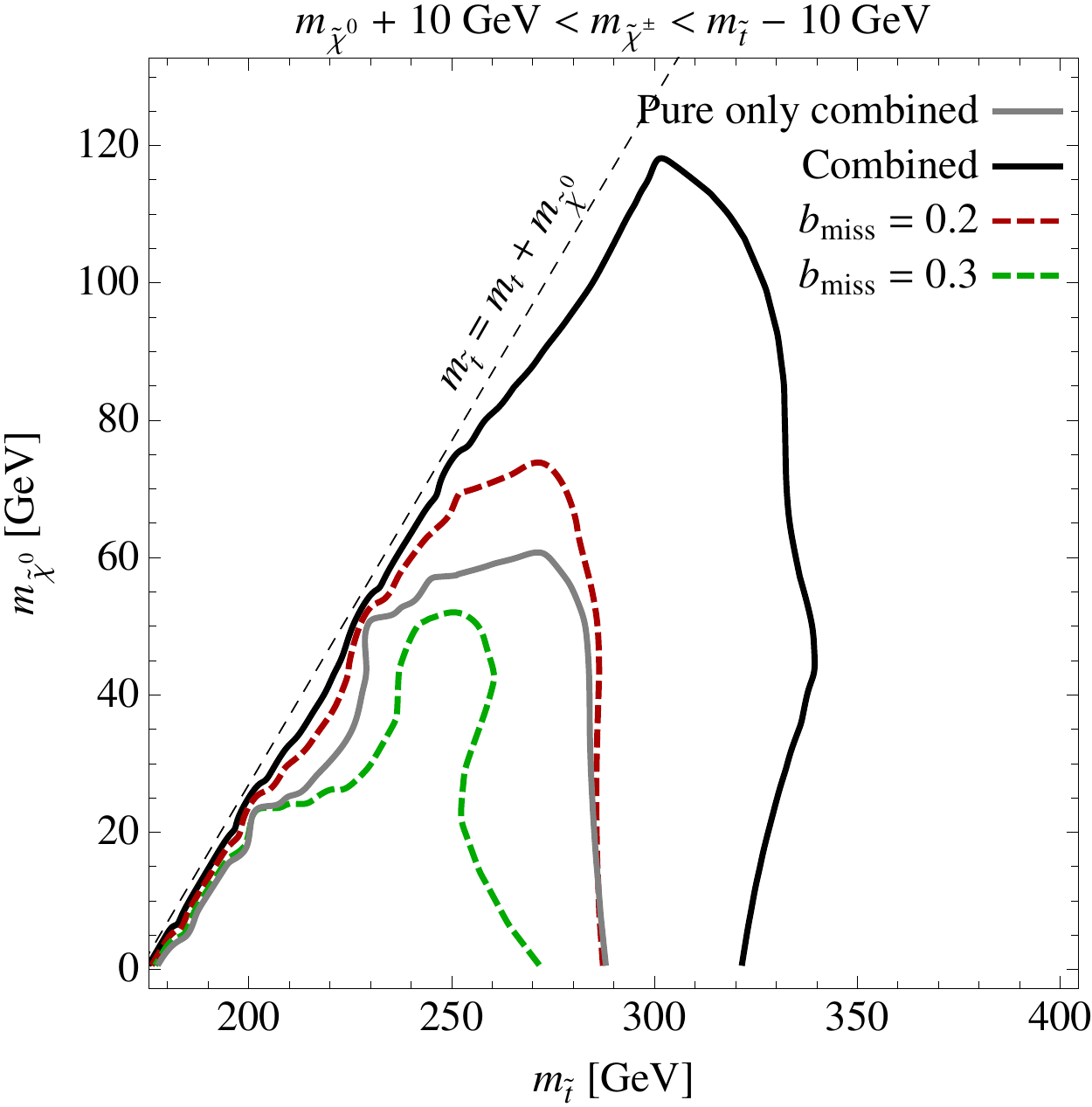}
\end{center}
\caption{A model independent combination of the 95\% $CL_s$ limits released by the ATLAS collaboration in refs.~\cite{ATLAS-CONF-2013-024, ATLAS-CONF-2013-037, ATLAS-CONF-2013-048, Aad:2013ija} valid for all choices of branching ratios into top-neutralino and bottom-chargino, and all choices of chargino mass in the range $m_{\tilde{\chi}^0}+10\mbox{ GeV}<m_{\tilde{\chi}^\pm}<m_{\tilde{t}}-10\mbox{ GeV}$.  The limits for 100\% total branching ratio into either top-neutralino or bottom-chargino are shown as solid contours, both with and without constraints coming from mixed decays.  Limits for a non-zero branching ratio into other decay modes, $b_{\rm miss}$, always with constraints coming from mixed decays, are shown as dashed contours.\label{fig:cube}}
\end{figure}

For each value of stop and neutralino mass we consider chargino masses from $m_{\tilde{\chi}^\pm}=m_{\tilde{\chi}^0}+10$~GeV to $m_{\tilde{\chi}^\pm}=m_{\tilde{t}}-10$~GeV that are not already excluded by the LEP bound.  This ensures that both stop decay modes are on-shell and that no singular behaviour is encountered due to degeneracies in the mass spectrum.  At each point we simulate a simplified model to derive upper limits on the branching ratios for all three complete stop pair decay processes.  These constraints may be of use in future studies, so are provided as additional material with this paper.  We then define
\begin{align}
& B_{\rm pure}=\sqrt{B_{00}}+\sqrt{B_{\pm\pm}}+b_{\rm miss} \nonumber\\
& B_{\rm mixed}=2\max\sbrack{\sqrt{B_{00}},\sqrt{B_{\pm\pm}}}+b_{\rm miss}-\sqrt{(1-b_{\rm miss})^2-2B_{0\pm}} \nonumber\\
& B_{\rm comb}=\min\sbrack{B_{\rm pure},B_{\rm mixed}}
\end{align}
for a fixed branching ratio into missing final states, $b_{\rm miss}$.\footnote{$B_{\rm mixed}=2\max\sbrack{\sqrt{B_{00}},\sqrt{B_{\pm\pm}}}+b_{\rm miss}$ if $2B_{0\pm}>(1-b_{\rm miss})^2$}  If $B_{\rm comb}<1$, the conditions \eqref{eq:bb} are satisfied and the corresponding masses can be excluded for all possible choices of branching ratio.  For each value of stop and neutralino mass we find the chargino mass that yields the largest value of $B_{\rm comb}$.  If this remains less than one, the corresponding stop and neutralino mass values can be excluded for all possible choices of branching ratio \emph{and} all values of the chargino mass in the range considered (assuming that $B_{\rm comb}$ varies smoothly, which should be assured by the minimum mass splittings we have imposed).

Our results can be seen in figure \ref{fig:cube}.  The limits are weakened even further but, impressively, non-trivial limits on the stop and neutralino masses \emph{still} remain for $b_{\rm miss}=0$, with stops up to 340 GeV able to be ruled out.  While this sounds somewhat less impressive than the values of up to 700 GeV often claimed, we emphasise that these limits are valid for all possible choices of branching ratio in the top-neutralino/bottom-chargino system, and for all values of chargino mass in the range considered, so will be very difficult to avoid.  It is again clear that mixed decays play an extremely important role in setting these limits, providing further motivation for a dedicated analysis of such processes.

The utility of this fast limit on stop masses can be investigated by referring to the typical branching ratios encountered in natural SUSY\@.  A more general gaugino sector for an unpolarised, light stop can be considered by varying $\mu$, $M_1$ and $M_2$ uniformly between 100 and 1000 GeV, $\tan\beta$ uniformly between 2 and 20, and $m_{\tilde{t}}$ uniformly between 100 and 1000 GeV\@.  Calculating the total branching ratio of stops into states other than the lightest neutralino or chargino using {\tt SUSY-HIT 1.3}~\cite{Djouadi:2006bz}, we find that around 25\% of models have a branching ratio into missing final states of less than 5\%.

The limits seem less optimistic when stop decays into modes other than top-neutralino and bottom-chargino are allowed, i.e.\ when $b_{\rm miss}>0$.  However, we really have assumed the worst case scenario here.  In reality, many of the ``missing'' decays will proceed via decay chains involving multiple gauginos, so will have similar topologies to the simplified models already considered.  Thus the corresponding analyses will be at least partially sensitive to them, rendering our assumption of zero sensitivity far too strong.  Furthermore, when simplified models for other stop decay chains become available, it is straightforward to include them in these limits. 

\section{Discussion}
\label{sec:discussion}

Critics of simplified models point to the fact that the limits afforded by them are only really applicable to vanishingly small hypersurfaces in the large parameter space of sparticle masses and couplings, and hence are of no more use than the GUT scale models that were previously fashionable.  In practise, constraining a realistic SUSY model directly from simplified model results involves a great deal of work.

We have shown that one can, in fact, derive useful limits by combining simplified models in a simple fashion.  In particular, we have seen that:
\begin{itemize}
\item It is certainly true that a naive reading of mass limits from a single simplified model is far too optimistic.  This is, of course, obvious, since one can always change sparticle branching ratios such that the topology assumed in the simplified model will not occur.  It is nevertheless interesting to quantify the size of the effect.
\item By adding a simplified model with mixed decay processes for the stop quark, one can significantly improve the limits provided by simplified models, even without re-optimising analyses on the chosen model.
\item Decay modes for which there is no acceptance (or decay modes for which there are not simplified models available) can be handled in our procedure by defining the maximum branching ratio into the relevant final states.  Uncertainties in the stop pair production cross-section can be handled in a similar way.
\item The existing information from the ATLAS and CMS collaborations already allows non-trivial, model independent limits to be placed on the stop mass, for arbitrary branching ratios, without the need for detailed Monte Carlo simulation.
\end{itemize}
This information ultimately suggests that, by using a sensible choice of basis simplified models, one can maximise the limits afforded on sparticle properties.  This is not limited to direct stop production; a similar analysis of gluino pair production models would be particularly interesting.  Tools are also being developed to recast many other new physics scenarios in terms of simplified models \cite{Kraml:2013mwa}, further extending the applicability of our technique.

It might still be argued that our results require various assumptions on the sparticle mass spectrum, and the limits thus obtained are not truly model independent.  This is inevitable, at least until a complete basis of simplified models becomes available, but the method is still useful.  When performing global statistical fits, one typically has to evaluate millions of likelihoods at many points in a parameter space.  Performing a full Monte Carlo simulation at every point rapidly becomes unfeasible.  The method presented here allows one to reject many points almost immediately, simply from the calculated mass and decay spectrum and without the need for any further simulation.

\section{Conclusions}
\label{sec:conclusions}

Using simplified models by themselves is too simplistic an approach to provide limits on sparticle masses in realistic models.  But, when combined, simplified models provide a powerful tool for constructing limits that are robust, conservative and as model independent as possible.  In this paper we have developed a simple way of combining simplified models, that requires no additional computational simulation, and tested it by putting limits on the stop mass.  In particular, current ATLAS and CMS results exclude stop masses up to 340 GeV for neutralino masses up to 120 GeV in a model independent way, provided that the total branching ratio into channels other than top-neutralino and bottom-chargino is small, and that there is no mass difference smaller than 10 GeV in the mass spectrum.  Although the limits we find are considerably weaker than those released by the experimental collaborations using individual simplified models, and those released by collaborations using statistical fitting techniques, they are much more generally applicable as no detailed assumptions need to be made about the branching ratios in the model.

A key idea for obtaining non-trivial limits was to consider a simplified model for mixed decays \cite{Graesser:2012qy}, where pair produced stops each decay via a different decay mode.  Constraints on these processes allow for much stronger limits, even though a dedicated analysis has not yet been developed.  By commissioning such an analysis, the experimental collaborations will greatly increase their overall sensitivity to direct stop production.

While our technique does not require additional simulation to combine existing simplified models, we did perform additional simulation to extend the reach of the simplified models already studied by the experimental collaborations.  The results constrain the branching ratios for each complete stop pair decay process for a wide variety of stop, neutralino and chargino masses.  These constraints may well prove useful for future studies, so are included as additional material with this paper.

\subsubsection*{Acknowledgements}

We thank Tony Gherghetta and Til Eifert for helpful comments and discussion.  This work was supported by the Australian Research Council.

\appendix

\section{Simulation details}

Part of the work in this paper involves applying our simplified model combination technique to models that ATLAS have not yet considered.  For this reason, we need a method for determining the ATLAS constraints on generic models with direct stop production.  The method is as follows.

We use {\tt Pythia 8.176} \cite{Sjostrand:2006za,Sjostrand:2007gs,Desai:2011su} to generate 100 000 Monte Carlo events per point, before using {\tt Delphes 3.0.9} to simulate the ATLAS detector.  This includes custom modifications for applying absolute isolation cuts, and to reproduce the correct $b$-tagging efficiency.  Stop pair production cross-sections for each model are calculated at next-to-leading order using {\tt PROSPINO 2.1} \cite{Beenakker:1996ed}.

The ATLAS simulation setup differs due to the use of a much more advanced, GEANT4 detector simulation, and the use of Madgraph for signal processes when the initial state radiation becomes significant (e.g.\ models where stops decay to bottom-chargino, in which the bottom quark is soft and the leading jet results from initial state radiation).  In addition, the $CL_s$ limit setting procedure implemented by ATLAS has access to a wide range of experimental systematic uncertainties that we cannot consider here.  Nevertheless, we find that we can reproduce the released ATLAS limits rather well, provided that we introduce an extra systematic in our limit setting procedure that is tuned to reproduce the ATLAS results as closely as possible.

To reproduce each ATLAS analysis, we assume a simplified form for the likelihood in each signal region:
\begin{align}
\lambda & =\mu Ks(1+\delta_s\sigma_s)+b(1+\delta_b\sigma_b) \nonumber\\
-2\log\mathcal{L} & =-2(n\log\lambda-\lambda)+\delta_s^2+\delta_b^2
\end{align}
where $\lambda$ is the expected number of events in the signal region, $s$ is the expected number of signal events predicted by our simulations, $b$ is the expected number of background events provided by ATLAS, $\mu$ is the signal strength modifier (0 for the background-only hypothesis and 1 for the nominal signal hypothesis), $\delta_s$ and $\delta_b$ are the realised values of signal and background systematic error random variables (assumed to be normally distributed with expected value zero, and scaled by $\sigma_s$ and $\sigma_b$ so that their variance is unity), $K$ is a scaling parameter tuned to reproduce the released ATLAS limits for signal regions where the tuning of $\sigma_s$ and $\sigma_b$ was not sufficient (see below), and $n$ is the number of signal events observed by ATLAS in the signal region.  The systematic parameters and $K$ are taken to be flat over the simplified model parameter space for simplicity.

This likelihood is simply the product of a Poisson likelihood (for the observed number of events) and two normal likelihoods (for the systematic variables modelling the effect of control measurements). We profile out $\delta_s$ and $\delta_b$ as nuisance parameters. From the profile likelihood we construct the test statistic $q=-2\log(\mathcal{L}_{s+b}/\mathcal{L}_{b})$ (as described in \cite{Read:2002hq}), determine its distribution numerically for each set of parameter choices, and use this to compute $CL_s$ values to set our limits.

The widths of the distributions for the systematic variables are tuned so that our 95\% $CL_s$ limits on the number of non-Standard Model signal events in each signal region match the values reported by ATLAS\@.  For each analysis, the signal regions are then combined by taking the $CL_s$ limit at each model point from the signal region with the strongest expected limit (as predicted by our now-tuned model of the likelihood), and the resulting limits in the stop-neutralino-chargino parameter space are compared with those produced by ATLAS.\@  That is, for the slices of this parameter space in which ATLAS has provided limits.  In all but one case this procedure satisfactorily reproduced the ATLAS limits.

In the cases of the limits from the ATLAS 2 $b$-jet analysis \cite{Aad:2013ija}, the limits in the $m_{\tilde{\chi}^{\pm}}=150$ GeV and $\Delta m=m_{\tilde{\chi}^{\pm}}-m_{\tilde{\chi}^0}=20$ GeV planes did not immediately agree with those released by ATLAS\@.  In this case, the scaling parameter $K$ was tuned uniformly across all signal regions in the analysis until good agreement was reached; a value of $K=0.6$ was found to achieve this, except in the parameter regions where signal region B dominated the limit.  In this case, initial state radiation is important, which Pythia is not capable of simulating correctly.  Given that this signal region does not figure strongly in our final results, we therefore drop it from the analysis in the paper.

\begin{figure}[!t]
\begin{center}
\includegraphics[width=0.49\textwidth]{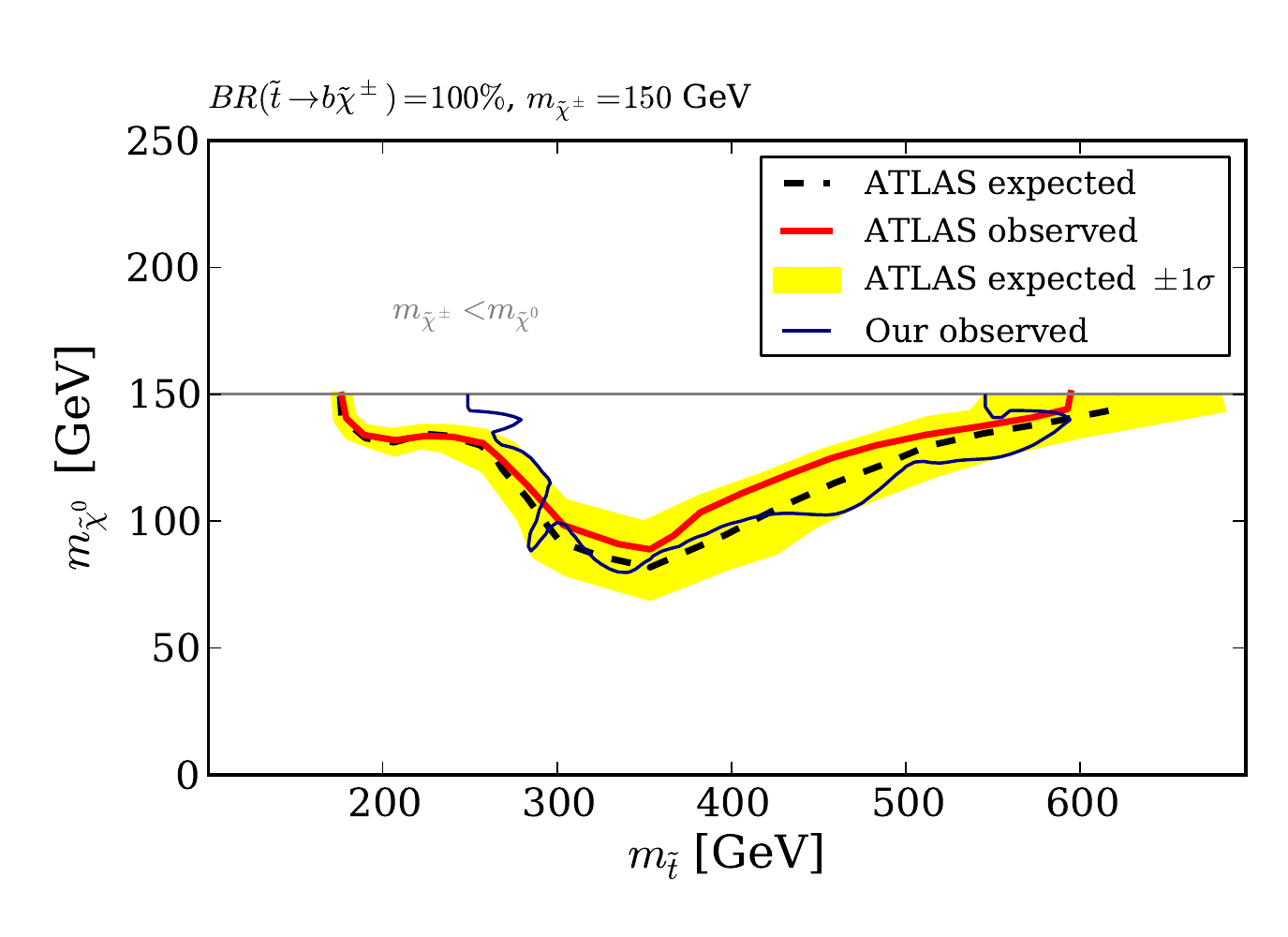}
\includegraphics[width=0.49\textwidth]{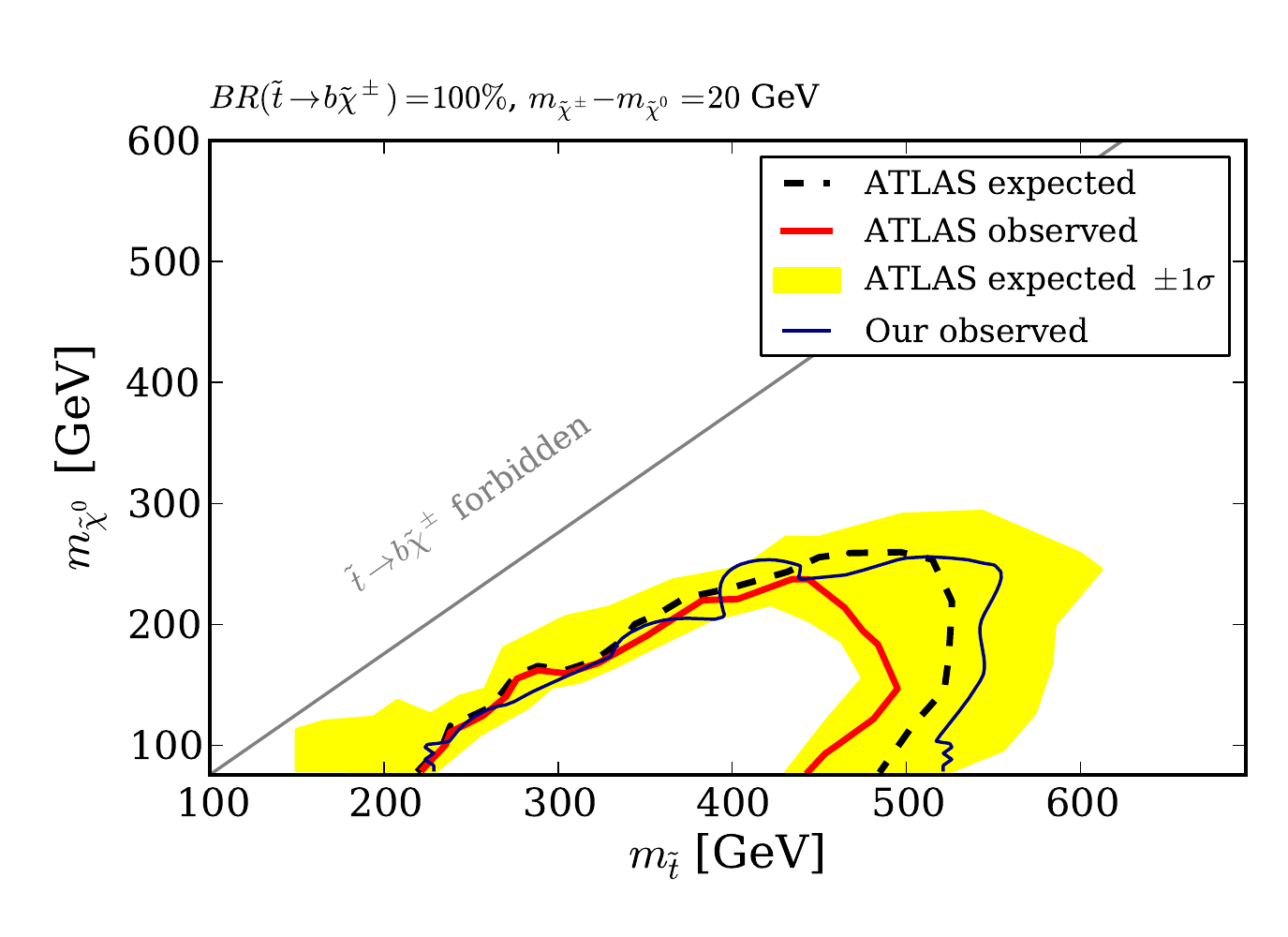}
\end{center}
\caption{95\% $CL_s$ limits from the ATLAS simplified model search for direct stop production in the 2 $b$-jet plus missing transverse energy final state \cite{Aad:2013ija}.  The limits on the stop and neutralino masses are shown for chargino masses of $m_{\tilde{\chi}^{\pm}}=150$ GeV (left) and $m_{\tilde{\chi}^{\pm}}=m_{\tilde{\chi}^0}+20$ GeV (right).  Agreement between our limits and the ATLAS limits at the level of the expected $1\sigma$ variation is observed everywhere, except in the $m_{\tilde{\chi}^{\pm}}=150$ GeV plane for stop masses below about 270 GeV, where signal region B (which we neglect) dominates.  Our limit is conservative in this region.\label{fig:USvsATLAS}}
\end{figure}

The limits we produce with signal region B removed are shown in figure \ref{fig:USvsATLAS}, superimposed on the official ATLAS limits.  This demonstrates agreement of the limits to within the expected statistical variation, except for a small region where signal region B dominates.  Here, we lose sensitivity and are left with a conservative limit.  With the full collection of likelihood models thus tuned, we are able to use them to produce the results of this paper.

\bibliographystyle{JHEP-2}
\bibliography{CSMbib}
\end{document}